\PassOptionsToPackage{unicode}{hyperref}
\PassOptionsToPackage{hyphens}{url}
\documentclass[
  11pt,
  openany]{article}
\usepackage{xcolor}
\usepackage[margin=1in]{geometry}
\usepackage{amsmath,amssymb}
\usepackage{iftex}
\ifPDFTeX
  \usepackage[T1]{fontenc}
  \usepackage[utf8]{inputenc}
  \usepackage{textcomp} % provide euro and other symbols
\else % if luatex or xetex
  \usepackage{unicode-math} % this also loads fontspec
  \defaultfontfeatures{Scale=MatchLowercase}
  \defaultfontfeatures[\rmfamily]{Ligatures=TeX,Scale=1}
\fi
\usepackage{lmodern}
\ifPDFTeX\else
\fi
\IfFileExists{upquote.sty}{\usepackage{upquote}}{}
\IfFileExists{microtype.sty}{% use microtype if available
  \usepackage[]{microtype}
  \UseMicrotypeSet[protrusion]{basicmath} % disable protrusion for tt fonts
}{}
\makeatletter
\@ifundefined{KOMAClassName}{% if non-KOMA class
  \IfFileExists{parskip.sty}{%
    \usepackage{parskip}
  }{% else
    \setlength{\parindent}{0pt}
    \setlength{\parskip}{6pt plus 2pt minus 1pt}}
}{% if KOMA class
  \KOMAoptions{parskip=half}}
\makeatother
\usepackage{color}
\usepackage{fancyvrb}

\DefineVerbatimEnvironment{Highlighting}{Verbatim}{commandchars=\\\{\}}
\newenvironment{Shaded}{}{}

\newcommand{\BuiltInTok}[1]{\textcolor[rgb]{0.00,0.50,0.00}{#1}}

\newcommand{\ControlFlowTok}[1]{\textcolor[rgb]{0.00,0.44,0.13}{\textbf{#1}}}

\newcommand{\DecValTok}[1]{\textcolor[rgb]{0.25,0.63,0.44}{#1}}

\newcommand{\FloatTok}[1]{\textcolor[rgb]{0.25,0.63,0.44}{#1}}

\newcommand{\KeywordTok}[1]{\textcolor[rgb]{0.00,0.44,0.13}{\textbf{#1}}}
\newcommand{\NormalTok}[1]{#1}
\newcommand{\OperatorTok}[1]{\textcolor[rgb]{0.40,0.40,0.40}{#1}}

\usepackage{longtable,booktabs,array}
\usepackage{calc} % for calculating minipage widths
\usepackage{etoolbox}
\makeatletter
\patchcmd\longtable{\par}{\if@noskipsec\mbox{}\fi\par}{}{}
\makeatother
\IfFileExists{footnotehyper.sty}{\usepackage{footnotehyper}}{\usepackage{footnote}}
\makesavenoteenv{longtable}
\usepackage{graphicx}
\makeatletter
\newsavebox\pandoc@box
\newcommand*\pandocbounded[1]{% scales image to fit in text height/width
  \sbox\pandoc@box{#1}%
  \Gscale@div\@tempa{\textheight}{\dimexpr\ht\pandoc@box+\dp\pandoc@box\relax}%
  \Gscale@div\@tempb{\linewidth}{\wd\pandoc@box}%
  \ifdim\@tempb\p@<\@tempa\p@\let\@tempa\@tempb\fi% select the smaller of both
  \ifdim\@tempa\p@<\p@\scalebox{\@tempa}{\usebox\pandoc@box}%
  \else\usebox{\pandoc@box}%
  \fi%
}
\def\fps@figure{htbp}
\makeatother
\providecommand{\tightlist}{%
  \setlength{\itemsep}{0pt}\setlength{\parskip}{0pt}}

\usepackage{longtable}
\usepackage{graphicx}
\usepackage{float}
\usepackage{array}
\usepackage{booktabs}
\usepackage{amsmath}
\usepackage{amssymb}
\AtBeginEnvironment{longtable}{\tiny\setlength{\tabcolsep}{2pt}}
\providecommand{\tightlist}{%
  \setlength{\itemsep}{0pt}\setlength{\parskip}{0pt}}

\renewcommand{\labelenumi}{\arabic{enumi}.}

\makeatletter
\def\fps@figure{H}
\makeatother
\usepackage{bookmark}
\IfFileExists{xurl.sty}{\usepackage{xurl}}{} % add URL line breaks if available
\hypersetup{
  pdftitle={Critical Volatility Threshold for Log-Normal to Power-Law Transition},
  pdfauthor={Valerii Kremnev},
  hidelinks,
  pdfcreator={LaTeX via pandoc}}

\title{Critical Volatility Threshold for Log-Normal to Power-Law
Transition}
\usepackage{etoolbox}
\makeatletter
\providecommand{\subtitle}[1]{% add subtitle to \maketitle
  \apptocmd{\@title}{\par {\large #1 \par}}{}{}
}
\makeatother
\subtitle{Iterated Options Model}
\author{Valerii Kremnev}
\date{January 2026}

\begin{document}
\maketitle

{
\setcounter{tocdepth}{2}
\tableofcontents
}
\section{Critical Volatility Threshold for Log-Normal to Power-Law
Transition: Iterated Options
Model}\label{critical-volatility-threshold-for-log-normal-to-power-law-transition-iterated-options-model}

\subsection{Abstract}\label{abstract}

Random walk models with log-normal outcomes fit local market
observations remarkably well. Yet interconnected or recursive structures
- layered derivatives, leveraged positions, iterative funding rounds -
periodically produce power-law distributed events. We show that the
transition from log-normal to power-law dynamics requires only three
conditions: randomness in the underlying process, rectification of
payouts, and iterative feed-forward of expected values. Using an
infinite option-on-option chain as an illustrative model, we derive a
critical volatility threshold at
\(\sigma^* = \sqrt{2\pi} \approx 250.66\%\) for the unconditional case.
With selective survival - where participants require minimum returns to
continue - the critical threshold drops discontinuously to
\(\sigma_{\text{th}}^{*} = \sqrt{\pi/2} \approx 125.3\%\), and can
decrease further with higher survival thresholds. The resulting outcomes
follow what we term the Critical Volatility (V*) Distribution - a
power-law whose exponent admits closed-form expression in terms of
survival pressure and conditional expected growth. The result suggests
that fat tails may be an emergent property of iterative log-normal
processes with selection rather than an exogenous feature.

\newpage

\subsection{Preface}\label{preface}

Financial systems are built on rectified payoffs. An investment in a
high-risk project returns either something or nothing - you cannot lose
more than you put in. An option pays \(\max(S - K, 0)\). Even limited
liability is a form of rectification.

These rectified structures often feed into one another. A successful
project enables others built on top of it. A successful trade becomes
the capital for the next trade. Derivative products reference other
derivative products. It would be useful to know how such iterations
behave - whether they remain stable or exhibit qualitatively different
dynamics.

To answer this, we analyze the limiting case: an infinite chain of
options, each written on the expected payout of the one before. The
result depends on three conditions - randomness in the underlying
process, rectification of payouts, and feed-forward of expected values.
These are sufficient to produce a critical threshold at
\(\sigma^* = \sqrt{2\pi} \approx 250.66\%\). Below this, cumulative
optionality remains bounded. Above it, the system diverges. With
selective survival - participants requiring minimum returns to continue
- the threshold drops to
\(\sigma_{\text{th}}^{*} = \sqrt{\pi/2} \approx 125.3\%\). The divergent
outcomes follow what we term the V* Distribution - a power-law whose
exponent depends on the specific volatility and participants'
willingness to make the next bet.

We also identify a self-similar regime at exactly the critical
threshold, where each iteration reproduces the statistical structure of
the previous one.

The conditions are minimal: randomness, bounded downside, and iteration.
These are not exotic assumptions, suggesting the mechanism may apply
broadly to dynamic systems with compounding behavior.

As for the infinite derivative tower itself: to our knowledge, no one
has built one. This is probably wise. But should financial engineering
continue its march toward increasingly layered products, at least the
location of the cliff is now known.

\newpage

\subsection{1. Introduction and
Motivation}\label{introduction-and-motivation}

The Black-Scholes framework provides a foundational model for pricing
European options. Under risk-neutral valuation, the price of a call
option reflects the expected value of its payoff \(\max(S_T - K, 0)\),
discounted appropriately. This ``rectification'' - the maximum of a
potentially negative quantity and zero - is the essential nonlinearity
that gives options their asymmetric payoff structure.

A natural question arises: what happens when we write an option on an
option? And then an option on that? In principle, one could construct an
arbitrarily deep tower of such instruments, each layer deriving its
value from the expected payout of the layer below.

This paper analyzes the mathematical structure of such iterated
rectified expectations. We find that:

\begin{enumerate}
\def\labelenumi{\arabic{enumi}.}
\item
  The system exhibits a phase transition at a critical volatility of
  \(\sigma^* = \sqrt{2\pi} \approx 250.66\%\) annualized for the
  unconditional ATM case. This assumes the perfect case, where pricing
  has no errors and volatility doesn't amplify between derivative layers
  but stays perfectly correlated to asset price at a constant ratio.
  Real systems, if built, will diverge much faster.
\item
  Below criticality (\(\sigma < \sigma^*\)), the total value of an
  infinite option chain converges to a finite sum, meaning optionality
  is ``bounded'' no matter how many layers are added.
\item
  Above criticality (\(\sigma > \sigma^*\)), the chain diverges - the
  cumulative value of optionality exceeds the underlying asset itself.
  This is not merely a mathematical curiosity; it implies that in
  extreme volatility regimes, the optionality can dominate the
  fundamental value of the product. This leads to amplification of
  expected payout at each consecutive step, making the expected payouts
  follow power-law dynamics.
\item
  At criticality (\(\sigma = \sigma^*\)), the system becomes
  self-similar, with each iteration reproducing the statistical
  structure of the previous one.
\item
  With selective survival - where participants require minimum returns
  to continue - the critical threshold drops to
  \(\sigma_{\text{th}}^{*} = \sqrt{\pi/2} \approx 125.3\%\). Power-law
  dynamics (the V* Distribution) emerge when \(\beta_{\text{eff}} > 1\),
  where \(\beta_{\text{eff}}\) is the conditional expected growth given
  survival.
\end{enumerate}

These findings have implications for understanding volatility regimes
during market stress, the pricing of compound options, and the
theoretical limits of derivative layering.

\newpage

\subsection{2. The Black-Scholes Setup}\label{the-black-scholes-setup}

\subsubsection{2.1 Standard Framework}\label{standard-framework}

Under the Black-Scholes model, the underlying asset follows geometric
Brownian motion:

\[dS_t = \mu S_t \, dt + \sigma S_t \, dW_t\]

The Black-Scholes formula for a European call option is:

\[C(S_t, t) = N(d_+) S_t - N(d_-) K e^{-r(T-t)}\]

where:

\[d_+ = \frac{1}{\sigma\sqrt{T-t}}\left[\ln\left(\frac{S_t}{K}\right) + \left(r + \frac{\sigma^2}{2}\right)(T-t)\right]\]

\[d_- = d_+ - \sigma\sqrt{T-t}\]

and \(N(\cdot)\) is the standard normal CDF.

\subsubsection{2.2 ATM Special Case}\label{atm-special-case}

For an at-the-money option where \(S_t = K\), we have
\(\ln(S_t/K) = 0\), so:

\[d_+ = \frac{r + \sigma^2/2}{\sigma}\sqrt{T-t}\]

\[d_- = \frac{r - \sigma^2/2}{\sigma}\sqrt{T-t}\]

In the \(r \ll \sigma\) limit (which holds for high-volatility regimes
where \(\sigma > 100\%\) and \(r \approx 5\%\)):

\[d_+ \approx \frac{\sigma^2/2}{\sigma}\sqrt{T-t} = \frac{\sigma}{2}\sqrt{T-t}\]

\[d_- \approx \frac{-\sigma^2/2}{\sigma}\sqrt{T-t} = -\frac{\sigma}{2}\sqrt{T-t}\]

Since \(d_- \approx -d_+\), we have \(N(d_-) = N(-d_+) = 1 - N(d_+)\),
and the option price simplifies to:

\[C = K\left[N(d_+) - N(d_-)\right] = K\left[2N\left(\frac{\sigma\sqrt{T-t}}{2}\right) - 1\right]\]

The ATM option price reduces to a single Gaussian CDF minus a constant.
Let us analyze its behavior.

\subsubsection{2.3 The Gaussian Structure}\label{the-gaussian-structure}

The Gaussian distribution appears explicitly in Black-Scholes through
\(N(d_+)\) and \(N(d_-)\), and simplifies to a single Gaussian CDF under
the high-volatility ATM assumption. We observe that the option payoff
\(\max(S_T - K, 0)\) cannot be negative by definition - the option
structure rectifies the underlying returns at zero.

The expected value of a rectified Gaussian is the mathematical core of
option pricing. We now analyze the general case of
\(\mathbb{E}[\max(X, 0)]\) for \(X \sim \mathcal{N}(\mu, \sigma^2)\).

In the following mathematical derivation, \(\sigma\) denotes the
standard deviation of the distribution
\(X \sim \mathcal{N}(\mu, \sigma^2)\), following standard statistical
convention, not the scale-less volatility.

\subsubsection{2.4 Rectified Gaussian
Expectations}\label{rectified-gaussian-expectations}

Let \(Y = \max(X, 0)\) where \(X \sim \mathcal{N}(\mu, \sigma^2)\). We
seek \(\mathbb{E}[Y]\).

The expectation splits into two regions:

\[\mathbb{E}[Y] = \mathbb{E}[X \cdot \mathbf{1}_{X > 0}] = \int_0^{\infty} x \cdot \frac{1}{\sigma\sqrt{2\pi}} e^{-(x-\mu)^2/2\sigma^2} dx\]

Substituting \(u = (x - \mu)/\sigma\):

\[\mathbb{E}[Y] = \int_{-\mu/\sigma}^{\infty} (\mu + \sigma u) \cdot \frac{1}{\sqrt{2\pi}} e^{-u^2/2} du\]

This separates into:

\[\mathbb{E}[Y] = \mu \int_{-\mu/\sigma}^{\infty} \phi(u) \, du + \sigma \int_{-\mu/\sigma}^{\infty} u \cdot \phi(u) \, du\]

The first integral is \(\mu \cdot \Phi(\mu/\sigma)\). For the second,
note that \(u \cdot \phi(u) = -\phi'(u)\), so:

\[\int_{-\mu/\sigma}^{\infty} u \cdot \phi(u) \, du = \left[-\phi(u)\right]_{-\mu/\sigma}^{\infty} = \phi(\mu/\sigma)\]

Therefore:

\[\mathbb{E}[Y] = \mu \Phi\left(\frac{\mu}{\sigma}\right) + \sigma \phi\left(\frac{\mu}{\sigma}\right)\]

where \(\Phi(\cdot)\) is the standard normal CDF and \(\phi(\cdot)\) is
the standard normal PDF.

\subsubsection{2.5 The Function g(z)}\label{the-function-gz}

Normalizing by \(\sigma\) and letting \(z = \mu/\sigma\):

\[\frac{\mathbb{E}[Y]}{\sigma} = z\Phi(z) + \phi(z)\]

Expanding using the integral forms of \(\Phi\) and \(\phi\):

\[\frac{\mathbb{E}[Y]}{\sigma} = \frac{1}{\sqrt{2\pi}}\left( z \int_{-\infty}^{z} e^{-t^2/2} dt + e^{-z^2/2} \right)\]

We define:

\[g(z) = z \int_{-\infty}^{z} e^{-t^2/2} dt + e^{-z^2/2}\]

so that the expected value of the rectified Gaussian becomes:

\[\mathbb{E}[Y] = \frac{\sigma}{\sqrt{2\pi}} g(z)\]

This form will be essential for analyzing iterations.

\newpage

\subsection{3. Iterated Options: The Mathematical
Structure}\label{iterated-options-the-mathematical-structure}

\subsubsection{3.1 The Iteration Scheme}\label{the-iteration-scheme}

Consider a chain of options where each option is written on the expected
payout of the previous one. In a simplified model of such chain, we
would be putting a new derivative instrument with the price of expected
payout of the previous one, and calculate new parameters. Since the
strike price from previous option would be just a constant multiplier,
we can focus on analyzing the rectified Gaussian behavior as a
simplified model.

Let \(\mu_n\) denote the expected value at stage \(n\), and suppose each
stage has volatility \(\sigma_n\).

From Section 2.5, the expected value of the rectified Gaussian at stage
\(n\) is:

\[\mathbb{E}_n[Y] = \frac{\sigma_n}{\sqrt{2\pi}} g(z_n)\]

where \(z_n = \mu_n/\sigma_n\).

If we let the output of one rectification become the mean of the next
(i.e., \(\mu_{n+1} = \mathbb{E}_n[Y]\)), we obtain:

\[\mu_{n+1} = \frac{\sigma_n}{\sqrt{2\pi}} g(z_n)\]

\subsubsection{3.2 General Recursion}\label{general-recursion}

From Section 2.5:

\[\mathbb{E}_1[Y] = \frac{\sigma_1}{\sqrt{2\pi}} g(z_1)\]

Let the output become the mean of the next stage:
\(\mu_2 = \mathbb{E}_1[Y]\).

The next price is:

\[z_2 = \frac{\mu_2}{\sigma_2} = \frac{\sigma_1}{\sigma_2} \cdot \frac{1}{\sqrt{2\pi}} g(z_1)\]

Let \(r = \sigma_1/\sigma_2\) and \(w = g(z_1)\). Then:

\[z_2 = \frac{r}{\sqrt{2\pi}} w\]

The next expectation:

\[\mathbb{E}_2[Y] = \frac{\sigma_2}{\sqrt{2\pi}} g(z_2)\]

More generally, letting \(w_n = g(z_n)\) and
\(\alpha = \frac{r}{\sqrt{2\pi}}\):

\[w_{n+1} = g(\alpha \cdot w_n)\]

Explicitly:

\[w_{n+1} = \alpha w_n \int_{-\infty}^{\alpha w_n} e^{-t^2/2} \, dt + e^{-\alpha^2 w_n^2/2}\]

This is a nonlinear recursion whose behavior depends critically on
\(\alpha\).

\subsubsection{3.3 The Self-Similar Case}\label{the-self-similar-case}

When \(\alpha = 1\) (equivalently, \(r = \sqrt{2\pi}\), i.e.,
\(\sigma_1 = \sigma_2 \sqrt{2\pi}\)), the recursion simplifies to:

\[w_{n+1} = g(w_n)\]

This is pure iteration of \(g\) - the process becomes self-similar. The
ratio:

\[\frac{w_n}{w_{n+1}} = \frac{w_n}{g(w_n)}\]

suggests that the sequence will have its own convergence/divergence
behavior depending on \(w_n\) (or equivalently \(\alpha\)). The
parameter \(\alpha\) controls how the recursion scales, determining
whether iterated expectations grow, shrink, or stabilize.

\newpage

\subsection{4. The Recentered (ATM) Case}\label{the-recentered-atm-case}

\subsubsection{4.1 Introducing the Shift}\label{introducing-the-shift}

In practice, options are often struck at-the-money (ATM), where the
strike equals the current expected value. We model this by introducing a
shift parameter \(s_n\) that recenters the distribution at each step:

\[z_n^s = \frac{\mu_n - s_n}{\sigma_n} = z_n - \frac{s_n}{\sigma_n}\]

Setting \(s_n = \mu_n\) (the ATM condition) forces \(z_n^s = 0\) at
every iteration.

The intuition for this shift is the following: each new option is
written ATM at inception, with strike equal to the current underlying
price (which is the expected value from the previous stage). The
underlying then fluctuates around this strike with its own volatility
over the holding period. Since the strike equals the mean, the
payoff-relevant distribution is centered at zero.

\subsubsection{4.2 Evaluation at Zero}\label{evaluation-at-zero}

Since \(g(0) = 0 + e^{0} = 1\), the shifted expectation simplifies
dramatically:

\[\mathbb{E}_n^s[Y] = \frac{\sigma_n}{\sqrt{2\pi}}\]

This is the well-known result that an ATM option's expected payout
(before discounting and without market price multiplier) is proportional
to volatility. It connects directly to the well-known practitioner's
approximation:

\[C \approx \frac{S \cdot \sigma\sqrt{T}}{\sqrt{2\pi}} \approx 0.4 \cdot S \cdot \sigma\sqrt{T}\]

\subsubsection{4.3 The Geometric Regime}\label{the-geometric-regime}

In financial contexts, we frequently assume that volatility is a
percentage related to the price - a stock with higher price has
proportionally higher absolute volatility. We use a similar definition
which scales with expected values.

We now return to the finance convention where \(\sigma\) denotes
percentage volatility (coefficient of variation, \(\sigma_n/\mu_n\)).
Assuming constant percentage volatility (absolute volatility scales
proportionally with price):

\[\mu_{n+1} = \frac{\sigma \cdot \mu_n}{\sqrt{2\pi}} = \beta \cdot \mu_n\]

where \(\beta = \sigma/\sqrt{2\pi}\).

This yields the closed form:

\[\mu_n = \mu_1 \cdot \beta^{n-1}\]

The expected value at each stage forms a geometric sequence.

\newpage

\subsection{5. Convergence and the Critical
Threshold}\label{convergence-and-the-critical-threshold}

\subsubsection{5.1 Sum of the Infinite
Chain}\label{sum-of-the-infinite-chain}

The total expected value across an infinite chain of ATM options is:

\[\sum_{n=1}^{\infty} \mu_n = \mu_1 \sum_{n=0}^{\infty} \beta^n = \frac{\mu_1}{1-\beta}\]

This converges if and only if \(\beta < 1\).

\subsubsection{5.2 The Critical
Volatility}\label{the-critical-volatility}

The convergence condition \(\beta < 1\) translates to:

\[\frac{\sigma}{\sqrt{2\pi}} < 1 \implies \sigma < \sqrt{2\pi} \approx 2.5066\]

In percentage terms, the critical volatility is
\(\sigma^* \approx 250.66\%\) annualized.

\subsubsection{5.3 Closed Form for the
Sum}\label{closed-form-for-the-sum}

When \(\sigma < \sqrt{2\pi}\):

\[\sum_{n=1}^{\infty} \mu_n = \frac{\mu_1 \sqrt{2\pi}}{\sqrt{2\pi} - \sigma}\]

\newpage

\subsection{6. Divergence and Power-Law Beyond the Critical
Threshold}\label{divergence-and-power-law-beyond-the-critical-threshold}

\subsubsection{6.1 Exponential Growth in the Supercritical
Regime}\label{exponential-growth-in-the-supercritical-regime}

When \(\sigma > \sqrt{2\pi}\), we have
\(\beta = \sigma/\sqrt{2\pi} > 1\), and the expected values grow
exponentially:

\[\mu_n = \mu_1 \cdot \beta^{n-1}\]

Each iteration amplifies the previous expected value. The total sum
diverges - there is no finite bound on cumulative optionality.

\subsubsection{6.2 Survival Condition and Power-Law
Emergence}\label{survival-condition-and-power-law-emergence}

On the real market, participants do not receive the expected value -
they receive a realized draw from the distribution. The ability of
players to continue playing depends on their outcomes and risk
tolerance.

We can take simplified option-like model which scales payoff with the
bet, and assume payoff is \(\max(X, 0)\) where
\(X \sim \mathcal{N}(0, \sigma w)\) - a zero-centered Gaussian.
Participants require a minimum return to justify continued risk-taking.
Let \(k_{\text{th}}\) be the threshold multiplier: participants survive
only if their payoff exceeds \(k_{\text{th}} \cdot w\).

Since the payoff must be positive and exceed the threshold, survival
requires \(X \geq k_{\text{th}} \cdot w\). Standardizing to
\(Z = X/(\sigma w)\) where \(Z \sim \mathcal{N}(0, 1)\):

\[P(\text{survive}) = P\left(Z \geq \frac{k_{\text{th}}}{\sigma}\right) = 1 - \Phi\left(\frac{k_{\text{th}}}{\sigma}\right)\]

For example, with \(\sigma = 3\) and \(k_{\text{th}} = 2.5\), we have
\(k_{\text{th}}/\sigma \approx 0.833\), giving \(p \approx 0.20\) (20\%
survival rate per round).

The probability of surviving \(n\) consecutive stages is:

\[P(\text{survive } n \text{ stages}) = p^n\]

Among survivors, the expected wealth multiplier per stage is the
\textbf{conditional expectation given survival}. For the zero-centered
truncated normal:

\[\beta_{\text{eff}} = \mathbb{E}\left[\frac{X}{w} \,\Big|\, X \geq k_{\text{th}} w\right] = \sigma \cdot \frac{\phi(k_{\text{th}}/\sigma)}{1 - \Phi(k_{\text{th}}/\sigma)} = \sigma \cdot \frac{\phi(k_{\text{th}}/\sigma)}{p}\]

where \(\phi\) is the standard normal PDF.

\textbf{The V* Critical Threshold.} The phase transition to power-law
behavior occurs when \(\beta_{\text{eff}} = 1\). Setting
\(z = k_{\text{th}}/\sigma\), this condition gives:

\[\sigma_{\text{critical}} = \frac{1 - \Phi(z)}{\phi(z)}\]

This is the inverse Mills ratio. As \(z \to 0^+\) (threshold approaching
zero from above):

\[\sigma_{\text{th}}^{*} = \lim_{z \to 0^+} \frac{1 - \Phi(z)}{\phi(z)} = \frac{0.5}{1/\sqrt{2\pi}} = \sqrt{\frac{\pi}{2}} \approx 1.253 \approx 125.3\%\]

This reveals a second critical constant: the moment any positive
survival threshold is introduced, the critical volatility drops from
\(\sigma^* = \sqrt{2\pi} \approx 250.66\%\) to
\(\sigma_{\text{th}}^{*} = \sqrt{\pi/2} \approx 125.3\%\). This happens
because filtering out participants with \(X < 0\) (approximately half
the population) doubles the conditional growth rate of survivors.

For higher thresholds (\(z > 0\)), the critical volatility decreases
further. The critical curve in \((\sigma, k_{\text{th}})\) space is
parameterized by:

\[\sigma_{\text{critical}}(z) = \frac{1 - \Phi(z)}{\phi(z)}, \quad k_{\text{th, critical}}(z) = z \cdot \frac{1 - \Phi(z)}{\phi(z)}\]

The number of surviving processes decays exponentially (\(p^n\)), but
when \(\beta_{\text{eff}} > 1\), the value of each survivor grows
exponentially (\(\beta_{\text{eff}}^n\)). This combination produces
power-law distributed outcomes.

\subsubsection{6.3 The V* Distribution (Critical Volatility
Distribution)}\label{the-v-distribution-critical-volatility-distribution}

If \(N\) independent processes start, after \(n\) iterations
approximately \(N \cdot p^n\) survive, each with value proportional to
\(\beta_{\text{eff}}^n\).

Setting \(v = \beta_{\text{eff}}^n\) (the value), we have
\(n = \log(v)/\log(\beta_{\text{eff}})\), so:

\[\text{Number with value} > v \propto p^{n} = p^{\log(v)/\log(\beta_{\text{eff}})} = v^{\log(p)/\log(\beta_{\text{eff}})}\]

This is a power-law with exponent:

\[\alpha = -\frac{\log(p)}{\log(\beta_{\text{eff}})}\]

Since \(p < 1\) (survival is not guaranteed) and
\(\beta_{\text{eff}} > 1\) (supercritical with conditional growth), we
have \(\alpha > 0\): a proper power-law tail.

The V* Distribution is thus:

\[P(V > v) \propto \left(1 - \Phi\left(\frac{k_{\text{th}}}{\sigma}\right)\right)^{\frac{\log v}{\log\left(\sigma \cdot \frac{\phi(k_{\text{th}}/\sigma)}{1 - \Phi(k_{\text{th}}/\sigma)}\right)}}\]

where:

\begin{itemize}
\tightlist
\item
  \(k_{\text{th}}\) is the threshold multiplier (minimum payoff as
  multiple of wealth)
\item
  \(\sigma\) is the volatility parameter
\item
  \(\phi\), \(\Phi\) are the standard normal PDF and CDF
\end{itemize}

This can be written as \(P(V > v) \propto v^{-\alpha}\) where
\(\alpha = -\log(p)/\log(\beta_{\text{eff}})\).

\subsubsection{6.4 At The Value: Extending to Negative
Thresholds}\label{at-the-value-extending-to-negative-thresholds}

The survival condition in Section 6.2 requires
\(X \geq k_{\text{th}} \cdot w\), where \(X\) is the underlying return
(before rectification) and \(k_{\text{th}}\) is the threshold
multiplier. While the payoff remains \(\max(X, 0)\), the survival
condition is evaluated on \(X\) itself.

For \(k_{\text{th}} > 0\), participants require positive returns above a
threshold - a natural constraint for investors seeking real gains. For
\(k_{\text{th}} = 0\), participants continue if \(X > 0\); the option
paid something.

Mathematically, nothing prevents \(k_{\text{th}} < 0\). This models a
different game: participants accept losses to their wealth to continue
playing. If \(X = -0.5w\), the option pays zero, but a participant with
\(k_{\text{th}} = -1\) survives - they absorb the loss from reserves and
enter the next round.

This is no longer option-like behavior. With negative thresholds,
participants have linear exposure to losses up to
\(|k_{\text{th}}| \cdot w\). We call this regime \textbf{At The Value
(ATV)}: participants commit to continue through adverse outcomes,
accepting wealth destruction for the chance to remain in the game.

The ATV regime characterizes patient capital:

\begin{itemize}
\tightlist
\item
  Venture funds that continue supporting portfolio companies through
  down rounds
\item
  Strategic investors with long time horizons
\item
  Any participant with reserves who values continuation over immediate
  returns
\end{itemize}

The phase diagrams that follow extend into this negative
\(k_{\text{th}}\) region, revealing how the critical boundary behaves
when participants tolerate losses.

\subsubsection{6.5 Phase Space
Characterization}\label{phase-space-characterization}

The V* Distribution exists in a two-dimensional parameter space
\((\sigma, k_{\text{th}})\). Figure 1 shows this space with the critical
boundary \(\beta_{\text{eff}} = 1\) separating two regimes:

\begin{itemize}
\item
  \textbf{Subcritical region} (upper-left, blue):
  \(\beta_{\text{eff}} < 1\). Conditional growth does not compensate for
  attrition. Outcomes are thin-tailed.
\item
  \textbf{Supercritical region} (lower-right, red/orange):
  \(\beta_{\text{eff}} > 1\). Conditional growth exceeds attrition.
  Outcomes follow the V* Distribution with power-law tails.
\end{itemize}

\begin{figure}
\centering
\pandocbounded{\includegraphics[keepaspectratio,alt={V* Phase Transition in (\textbackslash sigma, k\_\{\textbackslash text\{th\}\}) space. Color indicates conditional growth factor \textbackslash beta\_\{\textbackslash text\{eff\}\}. The critical boundary (solid curve) where \textbackslash beta\_\{\textbackslash text\{eff\}\} = 1 separates subcritical (thin-tailed) from supercritical (V* power-law) regimes. Vertical lines mark \textbackslash sigma\_\{\textbackslash text\{th\}\}\^{}\{*\} = \textbackslash sqrt\{\textbackslash pi/2\} and \textbackslash sigma\^{}* = \textbackslash sqrt\{2\textbackslash pi\}.}]{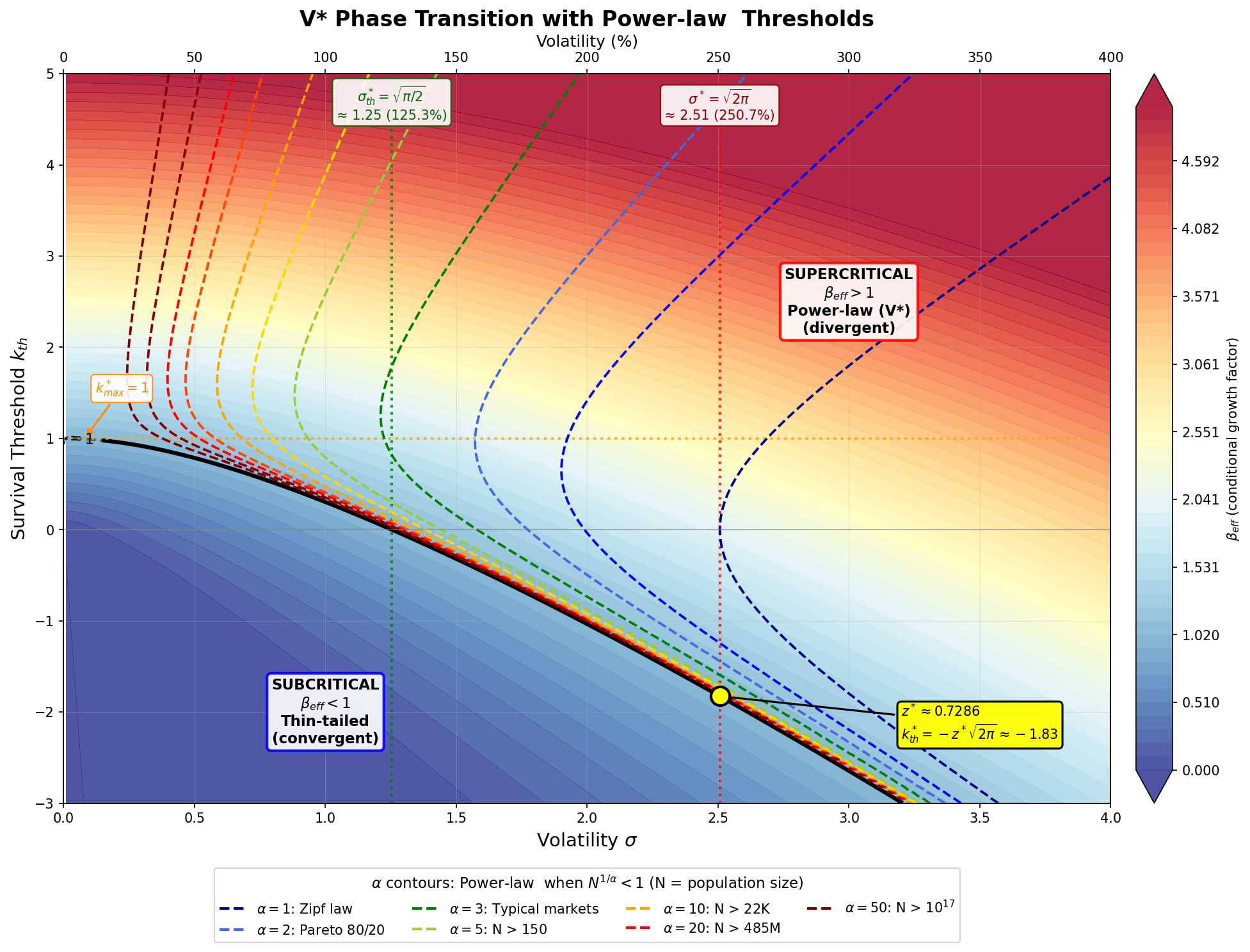}}
\caption{V* Phase Transition in \((\sigma, k_{\text{th}})\) space. Color
indicates conditional growth factor \(\beta_{\text{eff}}\). The critical
boundary (solid curve) where \(\beta_{\text{eff}} = 1\) separates
subcritical (thin-tailed) from supercritical (V* power-law) regimes.
Vertical lines mark \(\sigma_{\text{th}}^{*} = \sqrt{\pi/2}\) and
\(\sigma^* = \sqrt{2\pi}\).}
\end{figure}

Figure 2 decomposes the phase space into its constituent quantities. The
left panel shows survival probability
\(p = 1 - \Phi(k_{\text{th}}/\sigma)\), which decreases as the threshold
becomes more selective (higher \(k_{\text{th}}\)) or volatility
decreases (lower \(\sigma\)). The right panel shows the power-law
exponent \(\alpha = -\log(p)/\log(\beta_{\text{eff}})\) in the
supercritical region, with lower \(\alpha\) indicating heavier tails.

\begin{figure}
\centering
\pandocbounded{\includegraphics[keepaspectratio,alt={Phase space components. Left: Survival probability p per iteration. Right: V* power-law exponent \textbackslash alpha in the supercritical region. Lower \textbackslash alpha corresponds to heavier tails and more extreme wealth concentration.}]{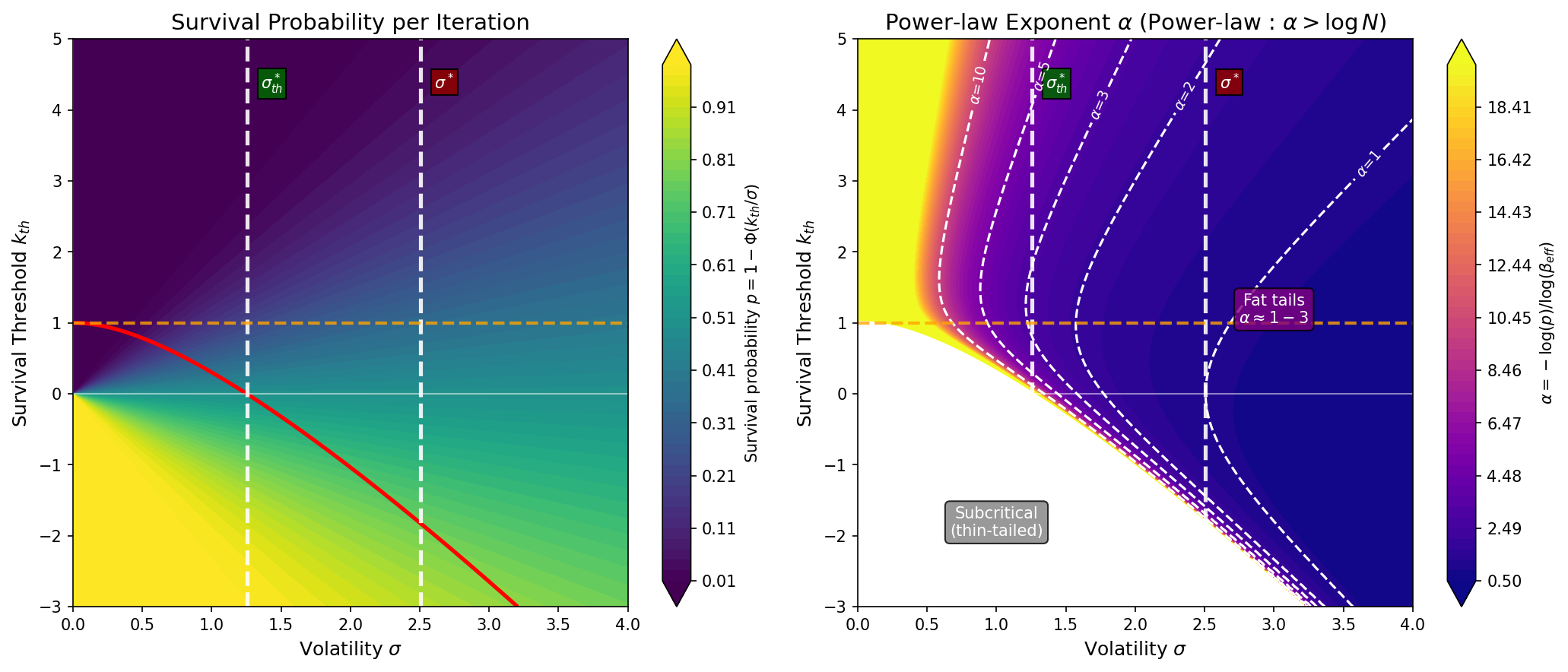}}
\caption{Phase space components. Left: Survival probability \(p\) per
iteration. Right: V* power-law exponent \(\alpha\) in the supercritical
region. Lower \(\alpha\) corresponds to heavier tails and more extreme
wealth concentration.}
\end{figure}

The shape of the critical boundary reflects a fundamental tradeoff. High
volatility generates large values among survivors; high selection
pressure concentrates growth among fewer participants. Both mechanisms
can produce \(\beta_{\text{eff}} > 1\):

\begin{itemize}
\tightlist
\item
  At low \(\sigma\), strong selection (high \(k_{\text{th}}\)) is
  required to achieve supercriticality
\item
  At high \(\sigma\), weaker selection (lower \(k_{\text{th}}\))
  suffices because volatility alone drives growth
\end{itemize}

The boundary curves through the parameter space accordingly, extending
into the risk-tolerant region (\(k_{\text{th}} < 0\)) at sufficiently
high volatility.

\subsubsection{6.6 Divergence versus Power-Law: The Role of
Selection}\label{divergence-versus-power-law-the-role-of-selection}

It is important to distinguish two phenomena that the framework reveals.

\textbf{Divergence of expected values} occurs when
\(\beta = \sigma/\sqrt{2\pi} > 1\), i.e., when \(\sigma > \sqrt{2\pi}\).
In this regime, the expected value at each iteration exceeds the
previous: \(\mu_{n+1} = \beta \cdot \mu_n\). The sum of an infinite
chain diverges. This is the result derived in Section 5 for the
unconditional ATM case.

\textbf{Power-law distribution} requires selection. The V* Distribution
emerges from the combination of two exponential processes:

\begin{itemize}
\tightlist
\item
  The number of surviving participants decays as \(p^n\)
\item
  The value of each survivor grows as \(\beta_{\text{eff}}^n\)
\end{itemize}

The power-law exponent \(\alpha = -\log(p)/\log(\beta_{\text{eff}})\) is
well-defined only when both \(p < 1\) (selection occurs) and
\(\beta_{\text{eff}} > 1\) (conditional growth exceeds unity). Without
selection (\(p = 1\)), there is no distribution of outcomes - all
participants follow the same trajectory.

These phenomena are related but distinct:

\begin{itemize}
\tightlist
\item
  Divergence concerns the total expected value of the system
\item
  Power-law concerns the shape of the outcome distribution under
  selection
\end{itemize}

The critical volatility \(\sigma^* = \sqrt{2\pi}\) marks where expected
values begin to diverge. But power-law behavior can emerge at any
volatility, provided selection is strong enough to achieve
\(\beta_{\text{eff}} > 1\). Conversely, even above \(\sigma^*\),
insufficient selection can fail to produce power-law tails if the
survival pool is too large.

\subsubsection{6.7 The Critical Intersection
Point}\label{the-critical-intersection-point}

The phase diagram reveals where these two thresholds intersect: the
point at which the unconditional divergence threshold
\(\sigma = \sqrt{2\pi}\) meets the critical boundary
\(\beta_{\text{eff}} = 1\).

At \(\sigma = \sqrt{2\pi}\), how much selection is required to produce
power-law behavior?

Setting \(\beta_{\text{eff}} = 1\):

\[\sigma \cdot \frac{\phi(z)}{1 - \Phi(z)} = 1 \quad \Rightarrow \quad \sigma = \frac{1 - \Phi(z)}{\phi(z)} = M(z)\]

where \(M(z)\) is the Mills ratio and \(z = k_{\text{th}}/\sigma\).
Substituting \(\sigma = \sqrt{2\pi}\):

\[M(z^*) = \sqrt{2\pi}\]

With \(\phi(z) = \frac{1}{\sqrt{2\pi}} e^{-z^2/2}\), this simplifies to:

\[1 - \Phi(z^*) = e^{-z^{*2}/2}\]

The solution is \(z^* \approx 0.7286\), corresponding to
\(k_{\text{th}}^* = -z^* \sqrt{2\pi} \approx -1.83\).

\textbf{Interpretation.} At \(\sigma = \sqrt{2\pi}\), the system
exhibits:

{\def\LTcaptype{none} % do not increment counter
\begin{longtable}[]{@{}
  >{\raggedright\arraybackslash}p{(\linewidth - 4\tabcolsep) * \real{0.3333}}
  >{\raggedright\arraybackslash}p{(\linewidth - 4\tabcolsep) * \real{0.3333}}
  >{\raggedright\arraybackslash}p{(\linewidth - 4\tabcolsep) * \real{0.3333}}@{}}
\toprule\noalign{}
{\bfseries 

Selection

} & {\bfseries 

Regime

} & {\bfseries 

Behavior

} \\
\midrule\noalign{}
\endhead
\bottomrule\noalign{}
\endlastfoot
\(k_{\text{th}} > -1.83\) & \(\beta_{\text{eff}} > 1\) & Power-law
(V*) \\
\(k_{\text{th}} = -1.83\) & \(\beta_{\text{eff}} = 1\) & Critical \\
\(k_{\text{th}} < -1.83\) & \(\beta_{\text{eff}} < 1\) & Thin-tailed \\
\end{longtable}
}

At the unconditional divergence threshold, expected values grow without
bound. Yet this divergence alone does not guarantee power-law outcomes.
If selection pressure is too weak (\(k_{\text{th}} < -1.83\)), the
survival pool includes too many participants, diluting conditional
growth below unity. The distribution remains thin-tailed despite
divergent expectations.

The constant \(z^*\) thus marks the minimum selection pressure required
to convert divergent growth into power-law distribution at
\(\sigma = \sqrt{2\pi}\). Above this point, selection concentrates
growth sufficiently for V* dynamics to emerge. Below it, dilution
dominates.

\subsubsection{6.8 Four Constants of Gaussian
Rectification}\label{four-constants-of-gaussian-rectification}

The framework yields four characteristic constants:

{\def\LTcaptype{none} % do not increment counter
\begin{longtable}[]{@{}
  >{\raggedright\arraybackslash}p{(\linewidth - 4\tabcolsep) * \real{0.3030}}
  >{\raggedright\arraybackslash}p{(\linewidth - 4\tabcolsep) * \real{0.2121}}
  >{\raggedright\arraybackslash}p{(\linewidth - 4\tabcolsep) * \real{0.4848}}@{}}
\toprule\noalign{}
\begin{minipage}[b]{\linewidth}\raggedright
{\bfseries 

Constant

}
\end{minipage} & \begin{minipage}[b]{\linewidth}\raggedright
{\bfseries 

Value

}
\end{minipage} & \begin{minipage}[b]{\linewidth}\raggedright
{\bfseries 

Interpretation

}
\end{minipage} \\
\midrule\noalign{}
\endhead
\bottomrule\noalign{}
\endlastfoot
\(\sigma^*\) & \(\sqrt{2\pi} \approx 2.507\) & Divergence threshold:
expected values unbounded \\
\(\sigma^*_{\text{th}}\) & \(\sqrt{\pi/2} \approx 1.253\) & Power-law
threshold at \(k_{\text{th}} \to 0^+\) \\
\(z^*\) & \(\approx 0.7286\) & Standardized selection threshold at
\(\sigma = \sigma^*\) \\
\(k^*_{\text{th}}\) & \(-z^*\sqrt{2\pi} \approx -1.83\) & Selection
threshold in parameter space \\
\end{longtable}
}

The ratio \(\sigma^*/\sigma^*_{\text{th}} = 2\) reflects the doubling of
conditional growth when the survival filter excludes negative outcomes.
The constant \(z^*\), defined by \(1 - \Phi(z^*) = e^{-z^{*2}/2}\), is
the standardized selection parameter at which power-law behavior first
emerges when volatility reaches the divergence threshold;
\(k^*_{\text{th}}\) expresses this in the units of the phase diagram.

These constants arise from the geometry of Gaussian rectification - the
interplay of tail probability, local density, and the normalization
factor \(\sqrt{2\pi}\).

\subsubsection{6.9 Implications}\label{implications}

The power-law exponent \(\alpha = -\log(p)/\log(\beta_{\text{eff}})\)
depends on both survival probability and conditional growth. Near
criticality (\(\beta_{\text{eff}} \approx 1\)), even modest selection
pressure produces heavy tails. Deep in the supercritical regime
(\(\beta_{\text{eff}} \gg 1\)), the distribution becomes increasingly
extreme - a few massive winners among many losers.

The phase diagrams reveal that V* dynamics are accessible across a wide
range of volatilities, provided selection is appropriately tuned.
Systems with moderate volatility (\(\sigma \approx 100-200\%\)) can
exhibit power-law behavior if participants impose sufficient selectivity
on continuation. Systems with extreme volatility can exhibit power-law
behavior even with weak selection.

This mechanism requires no exotic assumptions: iterated rectification of
a Gaussian process with selective continuation based on outcomes. The
fat tails emerge from the mathematics itself - specifically, from the
tension between exponential attrition and exponential conditional growth
that selection creates.

\newpage

\subsection{7. Numerical Simulations for
V*}\label{numerical-simulations-for-v}

To validate the theoretical predictions, we simulate a simplified model
which we call ATV (At The Value) where participants repeatedly bet their
entire wealth on an at-the-money option with payoff \(\max(X, 0)\) where
\(X \sim \mathcal{N}(0, \sigma w)\).

\subsubsection{7.1 Simulation Setup}\label{simulation-setup}

We simulate \(N = 10{,}000{,}000\) participants over \(T = 15\) periods,
each starting with wealth \(w_0 = \$20{,}000\). At each period,
participants in the high-risk game receive payoff \(\max(X, 0)\) where
\(X \sim \mathcal{N}(0, \sigma w)\). Participants drop out and switch to
a safe alternative (10\% volatility with in the money structure) if
their payoff falls below \(2.5 \times\) their current wealth -
representing the requirement that returns must justify continued
risk-taking in situations where bankruptcy risk is \textasciitilde50\%
per turn.

The model tests volatilities ranging from \(\sigma = 0.1\) (10\%) to
\(\sigma = 4.0\) (400\%), spanning the critical threshold at
\(\sigma^* = \sqrt{2\pi} \approx 2.507\) (251\%).

The simulation algorithm:

\small

\begin{Shaded}
\begin{Highlighting}[]
\KeywordTok{def}\NormalTok{ simulate\_atv\_model(n}\OperatorTok{=}\DecValTok{10\_000\_000}\NormalTok{, t}\OperatorTok{=}\DecValTok{15}\NormalTok{, w0}\OperatorTok{=}\DecValTok{20\_000}\NormalTok{, sigma}\OperatorTok{=}\FloatTok{2.5}\NormalTok{, threshold\_k}\OperatorTok{=}\FloatTok{2.5}\NormalTok{):}
\NormalTok{    w }\OperatorTok{=}\NormalTok{ np.full(n, w0, dtype}\OperatorTok{=}\BuiltInTok{float}\NormalTok{)}
\NormalTok{    in\_high\_risk }\OperatorTok{=}\NormalTok{ np.ones(n, dtype}\OperatorTok{=}\BuiltInTok{bool}\NormalTok{)}
\NormalTok{    sigma\_low }\OperatorTok{=} \FloatTok{0.1}

    \ControlFlowTok{for}\NormalTok{ year }\KeywordTok{in} \BuiltInTok{range}\NormalTok{(t):}
\NormalTok{        x\_high }\OperatorTok{=}\NormalTok{ np.random.randn(n) }\OperatorTok{*}\NormalTok{ sigma }\OperatorTok{*}\NormalTok{ w}
\NormalTok{        payoff\_high }\OperatorTok{=}\NormalTok{ np.maximum(x\_high, }\DecValTok{0}\NormalTok{)}

\NormalTok{        payoff\_low }\OperatorTok{=}\NormalTok{ w }\OperatorTok{*}\NormalTok{ (}\DecValTok{1} \OperatorTok{+}\NormalTok{ np.random.randn(n) }\OperatorTok{*}\NormalTok{ sigma\_low)}
\NormalTok{        payoff\_low }\OperatorTok{=}\NormalTok{ np.maximum(payoff\_low, }\DecValTok{0}\NormalTok{)}

\NormalTok{        threshold }\OperatorTok{=}\NormalTok{ threshold\_k }\OperatorTok{*}\NormalTok{ w}
\NormalTok{        dropout }\OperatorTok{=}\NormalTok{ in\_high\_risk }\OperatorTok{\&}\NormalTok{ (payoff\_high }\OperatorTok{\textless{}}\NormalTok{ threshold)}

\NormalTok{        w }\OperatorTok{=}\NormalTok{ np.where(in\_high\_risk, payoff\_high, payoff\_low)}
\NormalTok{        in\_high\_risk }\OperatorTok{=}\NormalTok{ in\_high\_risk }\OperatorTok{\&} \OperatorTok{\textasciitilde{}}\NormalTok{dropout}

    \ControlFlowTok{return}\NormalTok{ w}
\end{Highlighting}
\end{Shaded}

\normalsize

\subsubsection{7.2 Results}\label{results}

Figure 3 shows the rank-wealth distribution from simulation across all
volatility regimes on a log-log scale, with the V* theoretical
prediction overlaid (purple dashed line). The transition from curved
(log-normal) to linear (power-law) behavior is clearly visible as
volatility crosses the critical threshold. Figure 4 compares the wealth
distributions in subcritical and supercritical regimes, with the V*
theoretical power-law slope shown for comparison. Table 1 presents
detailed statistics for each volatility level.

\begin{figure}
\centering
\pandocbounded{\includegraphics[keepaspectratio,alt={Simulation vs V* Theory: Rank-wealth distribution across volatility regimes. Colored lines show simulation results from \textbackslash sigma=0.1 (blue) to \textbackslash sigma=4.0 (red). The purple dashed line shows the V* theoretical prediction (\textbackslash alpha = -\textbackslash log(p)/\textbackslash log(\textbackslash beta\_\{\textbackslash text\{eff\}\})) for \textbackslash sigma=3.0. Above criticality (\textbackslash sigma\^{}* \textbackslash approx 2.507), simulated distributions converge to the theoretical power-law slope.}]{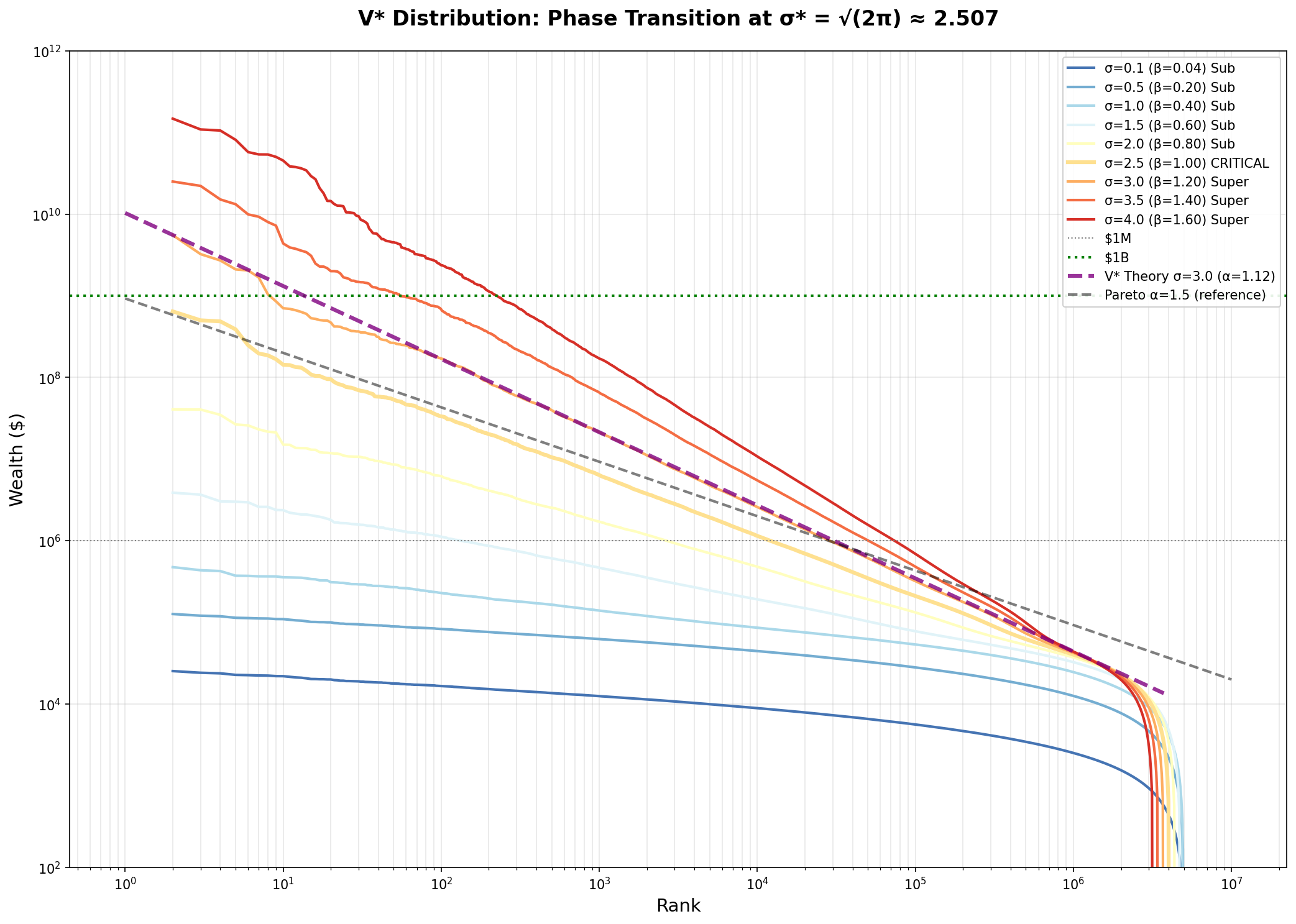}}
\caption{Simulation vs V* Theory: Rank-wealth distribution across
volatility regimes. Colored lines show simulation results from
\(\sigma\)=0.1 (blue) to \(\sigma\)=4.0 (red). The purple dashed line
shows the V* theoretical prediction
(\(\alpha = -\log(p)/\log(\beta_{\text{eff}})\)) for \(\sigma\)=3.0.
Above criticality (\(\sigma^* \approx 2.507\)), simulated distributions
converge to the theoretical power-law slope.}
\end{figure}

\begin{figure}
\centering
\pandocbounded{\includegraphics[keepaspectratio,alt={Simulation vs V* Theory: Subcritical (\textbackslash sigma=2.0, blue) vs supercritical (\textbackslash sigma=3.0, red) wealth distributions. Left: Probability density on log-log scale. Right: Rank-wealth plot with V* theoretical prediction (purple dashed) showing close agreement with supercritical simulation.}]{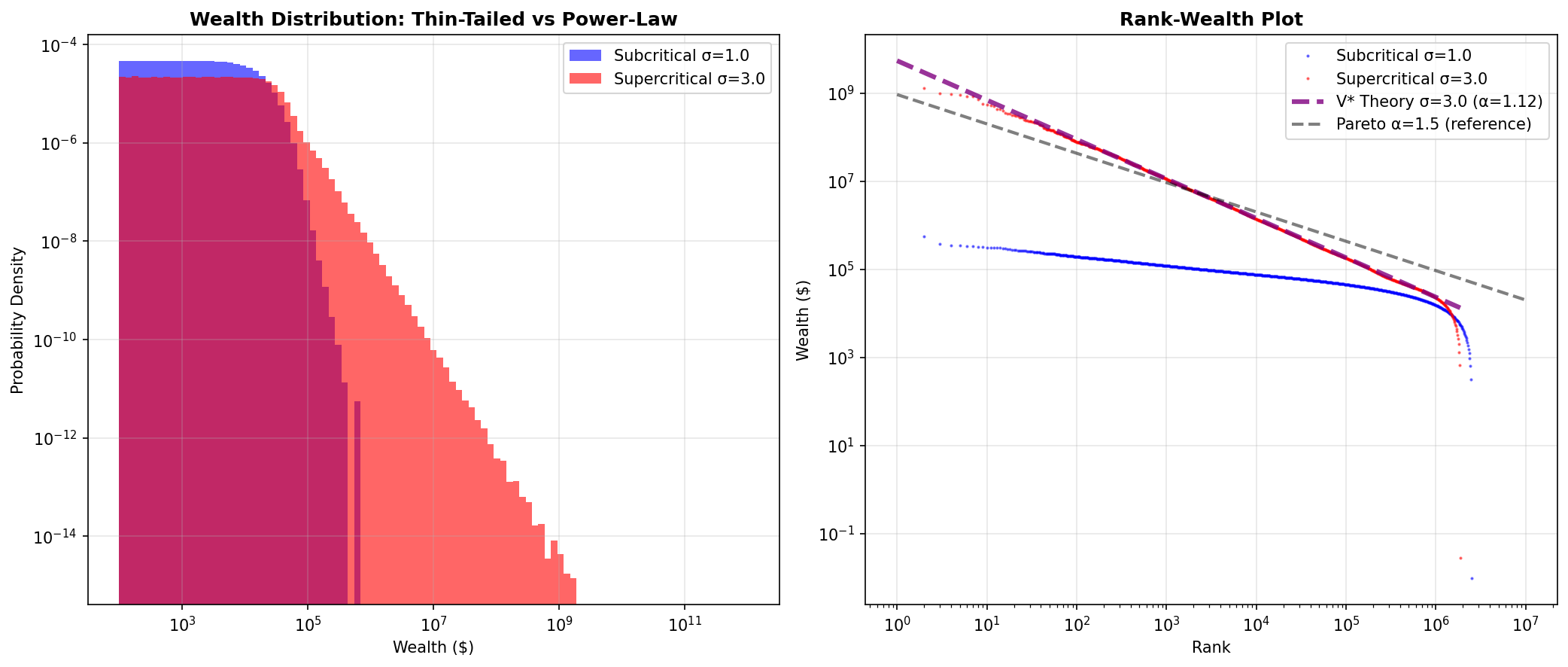}}
\caption{Simulation vs V* Theory: Subcritical (\(\sigma\)=2.0, blue) vs
supercritical (\(\sigma\)=3.0, red) wealth distributions. Left:
Probability density on log-log scale. Right: Rank-wealth plot with V*
theoretical prediction (purple dashed) showing close agreement with
supercritical simulation.}
\end{figure}

{\def\LTcaptype{none} % do not increment counter
\begin{longtable}[]{@{}
  >{\raggedright\arraybackslash}p{(\linewidth - 24\tabcolsep) * \real{0.0877}}
  >{\raggedright\arraybackslash}p{(\linewidth - 24\tabcolsep) * \real{0.0789}}
  >{\raggedright\arraybackslash}p{(\linewidth - 24\tabcolsep) * \real{0.0877}}
  >{\raggedright\arraybackslash}p{(\linewidth - 24\tabcolsep) * \real{0.1053}}
  >{\raggedright\arraybackslash}p{(\linewidth - 24\tabcolsep) * \real{0.0877}}
  >{\raggedright\arraybackslash}p{(\linewidth - 24\tabcolsep) * \real{0.0702}}
  >{\raggedright\arraybackslash}p{(\linewidth - 24\tabcolsep) * \real{0.0702}}
  >{\raggedright\arraybackslash}p{(\linewidth - 24\tabcolsep) * \real{0.0789}}
  >{\raggedright\arraybackslash}p{(\linewidth - 24\tabcolsep) * \real{0.0614}}
  >{\raggedright\arraybackslash}p{(\linewidth - 24\tabcolsep) * \real{0.0702}}
  >{\raggedright\arraybackslash}p{(\linewidth - 24\tabcolsep) * \real{0.0789}}
  >{\raggedright\arraybackslash}p{(\linewidth - 24\tabcolsep) * \real{0.0614}}
  >{\raggedright\arraybackslash}p{(\linewidth - 24\tabcolsep) * \real{0.0614}}@{}}
\toprule\noalign{}
\begin{minipage}[b]{\linewidth}\raggedright
{\bfseries 

\(\sigma\)

}
\end{minipage} & \begin{minipage}[b]{\linewidth}\raggedright
{\bfseries 

\(\beta\)

}
\end{minipage} & \begin{minipage}[b]{\linewidth}\raggedright
{\bfseries 

Bankrupt

}
\end{minipage} & \begin{minipage}[b]{\linewidth}\raggedright
{\bfseries 

Heavy Loss

}
\end{minipage} & \begin{minipage}[b]{\linewidth}\raggedright
{\bfseries 

\$2k-20k

}
\end{minipage} & \begin{minipage}[b]{\linewidth}\raggedright
{\bfseries 

\textgreater\$20k

}
\end{minipage} & \begin{minipage}[b]{\linewidth}\raggedright
{\bfseries 

\textgreater\$50k

}
\end{minipage} & \begin{minipage}[b]{\linewidth}\raggedright
{\bfseries 

\textgreater\$100k

}
\end{minipage} & \begin{minipage}[b]{\linewidth}\raggedright
{\bfseries 

\textgreater\$1M

}
\end{minipage} & \begin{minipage}[b]{\linewidth}\raggedright
{\bfseries 

\textgreater\$10M

}
\end{minipage} & \begin{minipage}[b]{\linewidth}\raggedright
{\bfseries 

\textgreater\$100M

}
\end{minipage} & \begin{minipage}[b]{\linewidth}\raggedright
{\bfseries 

\textgreater\$1B

}
\end{minipage} & \begin{minipage}[b]{\linewidth}\raggedright
{\bfseries 

Ratio

}
\end{minipage} \\
\midrule\noalign{}
\endhead
\bottomrule\noalign{}
\endlastfoot
0.10 & 0.04 & 5,229,613 & 3,313,805 & 1,456,562 & 20 & 0 & 0 & 0 & 0 & 0
& 0 & \(\infty\) \\
0.50 & 0.20 & 5,045,879 & 862,558 & 3,759,334 & 332,229 & 4,878 & 20 & 0
& 0 & 0 & 0 & \(\infty\) \\
1.00 & 0.40 & 5,054,141 & 436,565 & 3,093,118 & 1,416,176 & 134,222 &
4,702 & 0 & 0 & 0 & 0 & \(\infty\) \\
1.50 & 0.60 & 5,267,191 & 295,832 & 2,421,761 & 2,015,216 & 360,722 &
53,688 & 134 & 0 & 0 & 0 & \(\infty\) \\
2.00 & 0.80 & 5,603,627 & 225,395 & 1,960,344 & 2,210,634 & 562,968 &
158,609 & 2,672 & 34 & 1 & 0 & 78.6 \\
2.51 & 1.00 & 5,957,643 & 181,984 & 1,632,383 & 2,227,990 & 707,131 &
275,089 & 12,076 & 553 & 17 & 1 & 21.8 \\
3.00 & 1.20 & 6,277,439 & 153,424 & 1,398,570 & 2,170,567 & 794,856 &
371,871 & 28,760 & 2,234 & 175 & 8 & 12.9 \\
3.50 & 1.40 & 6,564,826 & 132,271 & 1,217,426 & 2,085,477 & 845,480 &
444,779 & 50,506 & 5,647 & 674 & 54 & 8.9 \\
4.00 & 1.60 & 6,818,493 & 116,154 & 1,075,393 & 1,989,960 & 870,073 &
497,072 & 73,912 & 10,648 & 1,602 & 224 & 6.9 \\
\end{longtable}
}

\textbf{Table 1:} Simulation results showing wealth distribution across
volatility regimes. ``Bankrupt'' = wealth \textless{} \$100, ``Heavy
Loss'' = \$100-\$2,000, ``Ratio'' = count(\$1M+) / count(\$10M+).

\subsubsection{7.3 Key Observations}\label{key-observations}

\textbf{The Low-Volatility Trap.} At \(\sigma = 0.1\), where
\(\beta = 0.04\), an astonishing 85\% of participants are either
bankrupt or suffer heavy losses. This occurs because the expected payoff
per period is only \(0.04w\) - a 96\% loss rate per iteration. Low
volatility provides insufficient upside to compensate for the inherent
bankruptcy risk of the ATV structure. Zero millionaires emerge from 10
million participants.

\textbf{Critical Transition.} At \(\sigma = 2.507 \approx \sigma^*\), we
observe \(\beta = 1.00\) - the break-even point where expected payoff
equals current wealth. The ratio of millionaires to decamillionaires
drops sharply to 21.8, indicating the emergence of heavy tails. The
first billionaire appears in the simulation.

\textbf{Supercritical Power-Law.} For \(\sigma > \sigma^*\), the ratio
stabilizes (21.8 → 12.9 → 8.9 → 6.9), the hallmark of power-law behavior
where the proportion between consecutive magnitude classes becomes
constant. At \(\sigma = 4.0\), despite 70\% bankruptcy or heavy loss,
224 billionaires emerge - a clear demonstration of the
few-massive-winners, many-losers distribution characteristic of power
laws.

\newpage

\subsection{8. Interpretation and Market
Implications}\label{interpretation-and-market-implications}

\subsubsection{8.1 Three Regimes}\label{three-regimes}

\textbf{Unconditional ATM Case.} When all participants continue
regardless of outcomes, the parameter \(\beta = \sigma/\sqrt{2\pi}\)
defines three regimes:

{\def\LTcaptype{none} % do not increment counter
\begin{longtable}[]{@{}
  >{\raggedright\arraybackslash}p{(\linewidth - 4\tabcolsep) * \real{0.2759}}
  >{\raggedright\arraybackslash}p{(\linewidth - 4\tabcolsep) * \real{0.3793}}
  >{\raggedright\arraybackslash}p{(\linewidth - 4\tabcolsep) * \real{0.3448}}@{}}
\toprule\noalign{}
\begin{minipage}[b]{\linewidth}\raggedright
{\bfseries 

Regime

}
\end{minipage} & \begin{minipage}[b]{\linewidth}\raggedright
{\bfseries 

Condition

}
\end{minipage} & \begin{minipage}[b]{\linewidth}\raggedright
{\bfseries 

Behavior

}
\end{minipage} \\
\midrule\noalign{}
\endhead
\bottomrule\noalign{}
\endlastfoot
\textbf{Subcritical} & \(\sigma < \sqrt{2\pi} \approx 250.66\%\) &
Convergent: Each payoff expected value is lower than previous and the
total payoff is bounded. Produces thin-tailed distribution of
outcomes. \\
\textbf{Critical} & \(\sigma \approx \sqrt{2\pi}\) & Self-similar: Each
layer reproduces the previous. \\
\textbf{Supercritical} & \(\sigma > \sqrt{2\pi}\) & Divergent: Each
payoff expected value is higher than previous and the total payoff goes
to infinity. \\
\end{longtable}
}

\textbf{V* Case with Survival Threshold.} When participants require
minimum returns \(k_{\text{th}}\) to continue, the conditional growth
factor
\(\beta_{\text{eff}} = \sigma \cdot \phi(k_{\text{th}}/\sigma)/p\)
determines the regime:

{\def\LTcaptype{none} % do not increment counter
\begin{longtable}[]{@{}
  >{\raggedright\arraybackslash}p{(\linewidth - 4\tabcolsep) * \real{0.2759}}
  >{\raggedright\arraybackslash}p{(\linewidth - 4\tabcolsep) * \real{0.3793}}
  >{\raggedright\arraybackslash}p{(\linewidth - 4\tabcolsep) * \real{0.3448}}@{}}
\toprule\noalign{}
\begin{minipage}[b]{\linewidth}\raggedright
{\bfseries 

Regime

}
\end{minipage} & \begin{minipage}[b]{\linewidth}\raggedright
{\bfseries 

Condition

}
\end{minipage} & \begin{minipage}[b]{\linewidth}\raggedright
{\bfseries 

Behavior

}
\end{minipage} \\
\midrule\noalign{}
\endhead
\bottomrule\noalign{}
\endlastfoot
\textbf{Subcritical} & \(\beta_{\text{eff}} < 1\) & Convergent:
Survivors do not grow fast enough to compensate for attrition. Produces
thin-tailed distribution. \\
\textbf{Critical} & \(\beta_{\text{eff}} \approx 1\) & Self-similar:
Conditional growth exactly balances survival probability. V* behavior
emerges. \\
\textbf{Supercritical} & \(\beta_{\text{eff}} > 1\) & Divergent:
Survivors grow faster than the population decays. Produces V*
Distribution with exponent
\(\alpha = -\log(p)/\log(\beta_{\text{eff}})\). \\
\end{longtable}
}

The V* framework generalizes the ATM result: with any positive survival
threshold, the critical volatility drops to
\(\sigma_{\text{th}}^{*} = \sqrt{\pi/2} \approx 125.3\%\), and decreases
further as \(k_{\text{th}}\) increases. Power-law dynamics can thus
emerge at volatilities far below \(\sigma^*\).

\subsubsection{8.2 Volatility of Options on
Options}\label{volatility-of-options-on-options}

An important real-world consideration: the volatility of an option's
value is generally \emph{higher} than the volatility of the underlying.
This is due to the convexity (gamma) of the option payoff. For a
compound option (option on an option), this effect compounds.

If we denote the volatility of the \(n\)-th layer as \(\sigma_n\),
empirically we observe:

\[\sigma_n > \sigma_{n-1}\]

This means that in practice, iterated option structures tend to
\emph{accelerate} toward the supercritical regime. The
constant-percentage-volatility assumption in our geometric regime is
thus conservative; real compound structures may diverge faster than our
model predicts.

\subsubsection{8.3 Time to Criticality}\label{time-to-criticality}

In Black-Scholes, the relevant volatility parameter is
\(\sigma\sqrt{T}\), where \(\sigma\) is the annualized volatility and
\(T\) is time to expiration in years. For the unconditional case, the
critical threshold \(\sigma\sqrt{T} = \sqrt{2\pi}\) gives:

\[T^* = \frac{2\pi}{\sigma^2}\]

For the V* case with survival threshold, the critical threshold drops to
\(\sigma\sqrt{T} = \sqrt{\pi/2}\), giving:

\[T_{\text{th}}^* = \frac{\pi/2}{\sigma^2}\]

{\def\LTcaptype{none} % do not increment counter
\begin{longtable}[]{@{}
  >{\raggedright\arraybackslash}p{(\linewidth - 4\tabcolsep) * \real{0.3077}}
  >{\raggedright\arraybackslash}p{(\linewidth - 4\tabcolsep) * \real{0.3462}}
  >{\raggedright\arraybackslash}p{(\linewidth - 4\tabcolsep) * \real{0.3462}}@{}}
\toprule\noalign{}
\begin{minipage}[b]{\linewidth}\raggedright
{\bfseries 

Annualized Vol \(\sigma\)

}
\end{minipage} & \begin{minipage}[b]{\linewidth}\raggedright
{\bfseries 

\(T^*\) (unconditional)

}
\end{minipage} & \begin{minipage}[b]{\linewidth}\raggedright
{\bfseries 

\(T_{\text{th}}^*\) (V* with threshold)

}
\end{minipage} \\
\midrule\noalign{}
\endhead
\bottomrule\noalign{}
\endlastfoot
10\% & 628 years & 157 years \\
20\% & 157 years & 39 years \\
50\% & 25 years & 6.3 years \\
100\% & 6.3 years & 1.6 years \\
150\% & 2.8 years & 8.4 months \\
200\% & 1.6 years & 4.7 months \\
250\% & 1.0 year & 3.0 months \\
300\% & 8.4 months & 2.1 months \\
400\% & 4.7 months & 1.2 months \\
500\% & 3.0 months & 3.3 weeks \\
800\% & 1.2 months & 1.3 weeks \\
\end{longtable}
}

For typical equity volatilities (15-30\%), the unconditional critical
threshold is centuries away. But with survival thresholds - which are
ubiquitous in real markets - criticality arrives four times faster. For
meme stocks and distressed names exhibiting 400-800\% implied
volatility, \textbf{V* criticality occurs within weeks}.

This means a 3-month ATM option on a 500\% vol underlying is already at
the critical regime - its expected payoff structure exhibits the
self-similar properties described in Section 3.3. A 6-month option on
the same underlying is supercritical.

\textbf{An interesting observation:} Early-stage startups exhibit annual
valuation volatility in the 100-250\% range, with funding rounds
occurring every 12-24 months. Under the unconditional model, the
critical horizon is 1-6 years - placing startups below criticality for
typical funding cycles. However, with survival thresholds
(\(\sigma_{\text{th}}^{*} \approx 125\%\)), the critical horizon drops
to 3-19 months - squarely within typical funding cycles.

Venture capitalists impose implicit survival thresholds: startups must
demonstrate sufficient progress to secure the next funding round. This
selective continuation - where only companies exceeding some return
threshold \(k_{\text{th}}\) survive to the next stage - creates
conditional growth \(\beta_{\text{eff}} > 1\) even when unconditional
\(\beta < 1\). The famously power-law distributed VC returns may thus be
a natural consequence of the V* mechanism: iterated optionality with
selective survival, where each funding stage represents both a survival
filter and a growth multiplier for those who pass.

\subsubsection{8.4 Connection to Real
Instruments}\label{connection-to-real-instruments}

Several existing instruments exhibit related dynamics:

\begin{itemize}
\tightlist
\item
  \textbf{Compound options} (options on options): Used in corporate
  finance for staged investments and in FX markets.
\item
  \textbf{Volatility derivatives}: VIX options are options on a
  volatility index, which is itself derived from option prices - a form
  of second-order optionality.
\item
  \textbf{Leveraged ETFs}: Daily rebalancing creates path-dependent
  compounding effects related to iterated expectations.
\item
  \textbf{Convertible bonds with call provisions}: Multiple embedded
  options create layered optionality.
\end{itemize}

Instruments involving averaging over multiple options (such as VIX
derivatives) present an interesting direction for potential extension of
the framework. The aggregation may produce lower volatility compared to
individual instruments, which warrants additional modelling not covered
in this paper.

\newpage

\subsection{9. Conclusion}\label{conclusion}

We have analyzed the behavior of iterated rectified Gaussian
expectations, illustrated by the theoretical construct of an infinite
chain of options-on-options. Our main findings:

\begin{enumerate}
\def\labelenumi{\arabic{enumi}.}
\item
  A critical volatility threshold exists at
  \(\sigma^* = \sqrt{2\pi} \approx 250.66\%\). Below this threshold, the
  cumulative value of an infinite option chain converges; above it, the
  chain diverges. This is the upper bound of stability under the
  idealized assumption of the system being maximally stable.
\item
  The supercritical regime implies that optionality can exceed
  underlying value. This is practically relevant during market stress
  events when implied volatilities spike above 250\%.
\item
  Real compound structures tend toward supercriticality because option
  volatility exceeds underlying volatility due to convexity effects.
  Real-world volatility amplification, leverage, or imperfect pricing
  would result in a lower critical bound.
\item
  With selective survival, the critical threshold drops discontinuously
  to \(\sigma_{\text{th}}^{*} = \sqrt{\pi/2} \approx 125.3\%\). We term
  the resulting power-law the V* Distribution, characterized by survival
  probability \(p = 1 - \Phi(k_{\text{th}}/\sigma)\) and conditional
  expected growth
  \(\beta_{\text{eff}} = \sigma \cdot \phi(k_{\text{th}}/\sigma)/p\).
  The power-law exponent \(\alpha = -\log(p)/\log(\beta_{\text{eff}})\)
  admits closed-form expression.
\end{enumerate}

The thresholds \(\sigma^* = \sqrt{2\pi}\) and
\(\sigma_{\text{th}}^{*} = \sqrt{\pi/2}\) emerge purely from the
geometry of Gaussian rectification - non-obvious boundaries that
separate fundamentally different economic regimes.

The V* Distribution provides a mechanism for power-law emergence that
requires no exotic assumptions. In repeated games with rectified
Gaussian payoffs, the number of surviving participants decays
exponentially as \(p^n\), while the wealth of each survivor grows
exponentially as \(\beta_{\text{eff}}^n\). The conditional nature of
\(\beta_{\text{eff}}\) is essential: it measures the expected growth
\emph{given survival}, not the unconditional expected payoff. This
interplay between exponential attrition and exponential conditional
growth produces fat tails with predictable exponents.

\textbf{A final observation:} While literal towers of
derivatives-on-derivatives are rare, our mathematical framework requires
a much weaker assumption - merely that expected returns propagate
through some iterative structure. Many common financial arrangements
satisfy this condition without being explicit derivative chains: loans
and credit facilities (where the borrower's ability to repay depends on
asset values), margin accounts and leveraged positions (where
maintenance requirements create recursive dependencies), and tightly
coupled instrument prices (where one instrument's value serves as
collateral or reference for another). The interaction of these
structures during stress events - when correlations spike and
volatilities exceed normal ranges - may exhibit dynamics similar to
those analyzed here, even without any formal options being written.

Furthermore, the framework does not require that all layers of the
derivative structure exist simultaneously. Consecutive dependent
instruments unfolding over time - where each stage's payout becomes the
underlying for the next - satisfy the same mathematical recursion. As we
have shown in simulation, the V* Distribution emerges reliably from this
process, with power-law exponents matching theoretical predictions. The
critical threshold we identify may thus be relevant not just for exotic
derivatives, but for understanding systemic behavior in leveraged,
interconnected financial systems evolving through time.

One might wonder if this model could help in predicting instability in
less exotic cases. Could black swans, fat tails, unexpected VC returns,
and volatility smiles have been predicted by feeding the random walk
back into itself and checking if it converges?

As a last note, we would like to emphasize that whether anyone should
actually construct an infinite derivative tower remains, we maintain,
inadvisable. But at least we now know where it would break.

\newpage

\subsection{Appendix: Notation Summary}\label{appendix-notation-summary}

{\def\LTcaptype{none} % do not increment counter
\begin{longtable}[]{@{}
  >{\raggedright\arraybackslash}p{(\linewidth - 2\tabcolsep) * \real{0.4000}}
  >{\raggedright\arraybackslash}p{(\linewidth - 2\tabcolsep) * \real{0.6000}}@{}}
\toprule\noalign{}
\begin{minipage}[b]{\linewidth}\raggedright
{\bfseries 

Symbol

}
\end{minipage} & \begin{minipage}[b]{\linewidth}\raggedright
{\bfseries 

Definition

}
\end{minipage} \\
\midrule\noalign{}
\endhead
\bottomrule\noalign{}
\endlastfoot
\(\Phi(z)\) & Standard normal CDF \\
\(\phi(z)\) & Standard normal PDF \\
\(g(z)\) & \(z \int_{-\infty}^{z} e^{-t^2/2} dt + e^{-z^2/2}\),
unnormalized expected rectified value \\
\(\sigma\) & Volatility parameter \\
\(\beta\) & \(\sigma/\sqrt{2\pi}\), unconditional geometric ratio \\
\(\sigma^*\) & \(\sqrt{2\pi} \approx 2.5066 \approx 250.66\%\), critical
volatility (unconditional) \\
\(\sigma_{\text{th}}^{*}\) &
\(\sqrt{\pi/2} \approx 1.2533 \approx 125.3\%\), critical volatility
(with any survival threshold) \\
\(k_{\text{th}}\) & Threshold multiplier (minimum payoff as multiple of
wealth) \\
\(p\) & Survival probability per stage,
\(1 - \Phi(k_{\text{th}}/\sigma)\) \\
\(\beta_{\text{eff}}\) & Conditional expected growth factor,
\(\sigma \cdot \phi(k_{\text{th}}/\sigma) / p\) \\
\(\alpha\) & Power-law exponent,
\(-\log(p)/\log(\beta_{\text{eff}})\) \\
\(V^*\) & V* Distribution: \(P(V > v) \propto v^{-\alpha}\) \\
\end{longtable}
}

\newpage

\subsection{Code Availability}\label{code-availability}

Simulation code and supplementary materials are available at:
\href{https://github.com/sci2sci-opensource/research/tree/master/critical-volatility-and-v*}{github.com/sci2sci-opensource/research}

\newpage

\subsection{References}\label{references}

Black, F., \& Scholes, M. (1973). The pricing of options and corporate
liabilities. \emph{Journal of Political Economy}, 81(3), 637-654.

Geske, R. (1979). The valuation of compound options. \emph{Journal of
Financial Economics}, 7(1), 63-81.

\end{document}